\documentstyle[12pt,epsfig]{article}

\newcommand{\be}{\begin{equation}}
\newcommand{\ee}{\end{equation}}
\begin{document}  
\topmargin 0pt
\oddsidemargin=-0.4truecm
\evensidemargin=-0.4truecm
\renewcommand{\thefootnote}{\fnsymbol{footnote}}
\newpage
\setcounter{page}{0}
\begin{titlepage}   
\vspace*{-2.0cm}  
\begin{flushright}
FISIST/13-2002/CFIF \\
hep-ph/0206193
\end{flushright}
\vspace*{0.1cm}
\begin{center}
{\Large \bf
Resonance Spin Flavour Precession of Solar Neutrinos After SNO NC Data}
\vspace{0.6cm}

\vspace{0.6cm}

{\large 
Bhag C. Chauhan\footnote{On leave from Govt. Degree College, Karsog (H P)
India - 171304. 
E-mail: chauhan@cfif.ist.utl.pt} and
Jo\~{a}o Pulido\footnote{E-mail: pulido@cfif.ist.utl.pt}}\\  
\vspace{0.15cm}
{\em Centro de F\'\i sica das Interac\c c\~oes Fundamentais (CFIF)} \\
{\em Departamento de F\'\i sica, Instituto Superior T\'ecnico }\\
{\em Av. Rovisco Pais, P-1049-001 Lisboa, Portugal}\\  
\end{center}
\vglue 0.6truecm
\begin{abstract}
We present an analysis of the solar neutrino data assuming the deficit of
solar neutrinos to be originated from the interaction of their transition
magnetic moments with the solar magnetic field. We perform fits to the rates
only and global fits and consider separately the existing data prior to the
announcement of the SNO NC results, and present data. Predictions for the 
Borexino experiment are also derived. The solar field profiles
are taken both in the radiation zone and core of the sun, and in the convective 
zone. The latter are chosen so as to exhibit a rapid increase across the bottom 
of the convective zone and a moderate decrease towards the surface. Regarding the
field profiles in the radiative zone and core, it is found that the data show
a preference for those cases in which a strong field rests at the solar centre
with a steep decrease thereafter. For these, the quality of the global fits is
as good as the one from the best oscillation solutions and the same as for the
convective zone profiles examined. It is also found that the $\chi^2$ of the
fits increases when the most recent data are considered, owing to the 
smaller errors involved. This in turn provides more precise predictions for
Borexino than previous ones, thus resulting in a clearer possible distinction 
between magnetic moment and the currently favoured oscillation solutions.   
\end{abstract}
\vspace{.5cm}
\centerline{Pacs numbers: 14.60.Pq, 26.65+t}
\centerline{Keywords: neutrino magnetic moment, solar neutrino problem}
\end{titlepage}   
\renewcommand{\thefootnote}{\arabic{footnote}}
\setcounter{footnote}{0}
\section{Introduction}
If neutrinos have a sizeable magnetic moment \cite{Cisneros} 
their interaction with the solar magnetic field can turn active 
${\nu_e}_L$ 's produced in the core of the sun into right handed 
antineutrinos of a different flavour or into sterile neutrinos, unseen 
in terrestrial experiments. This precession can be resonantly enhanced 
in matter \cite{LMA} with the location of the critical density being 
determined by the neutrino energy, in much the same way as the resonant 
amplification of oscillations, the MSW mechanism \cite{MSW}. For Majorana 
neutrinos only transition moments are possible and the interaction causes 
a simultaneous flip of spin and flavour, so that the resulting antineutrino 
can still be detected in neutrino electron scattering experiments, while
in the Dirac case the final state remains undetectable. 

The resonance spin flavour precession of solar neutrinos (RSFP) has not 
received as much attention as oscillations, possibly due to the fact that 
it requires a large neutrino magnetic moment $O(10^{12}-10^{-11})\mu_B$, 
far beyond the electroweak standard model value. Nevertheless, several
analyses exist \cite{GN,PA,AP,MR,Gago} which show that RSFP provides 
excellent fits to solar neutrino data, in some cases better than the best 
oscillation solution, the LMA one. While the much expected Kamland results 
\cite{Kamland} are unavailable and all possibilities remain open, 
it is very important to test 'non-standard'
solar neutrino solutions, of which RSFP is the most plausible one.
Furthermore RSFP has the interesting feature of 
providing a close relationship between the energy shape of the survival
probability and the solar magnetic field profile, in the sense that the 
most suppressed neutrinos have their resonance located in the region 
where the field is the strongest \cite{PRD}.   

In this paper we present an investigation of all the solar neutrino data 
including the recent SNO results on the charged current day/night effect 
and the neutral current, based on the assumption that neutrinos undergo 
RSFP inside the sun. We neglect the possible contribution of flavour
mixing, that is we take the angle $\theta_{12}$ to be too small
to play any role in the solar neutrino problem. Only Majorana neutrinos 
are considered, not only because these have been known for some time 
to provide better fits to solar data than Dirac ones \cite{GN}, 
\footnote{In fact the early comparisons between Kamiokande and Chlorine 
data alone, with Kamiokande showing a larger signal than Chlorine, always 
favoured the possibility that non ${\nu_e}_L$' s were active.} but 
mainly because the new SNO NC data seem to exclude the latter if RSFP 
happens to be the solution.  

The available information on the solar magnetic field is still quite 
limited at present \cite{ACT} and some authors \cite{FG} argue that a large
field in the convective zone may not be possible, since it would show 
up as an 11 year cycle in the SuperKamiokande \cite{SK} data which is
known not to be the case. Instead they consider a large field in the 
lower radiative zone and the core where most neutrinos are produced. It 
remains unclear however whether the sunspot cycle effect extends all 
the way down to the bottom of the convective zone. Hence other authors
\cite{ACT}, \cite{MR} favour a profile exhibiting a peak at the bottom
of the convective zone with a moderate decrease up to the surface where
it nearly vanishes. In our present analyses we consider profiles both
in the radiative zone and core, and in the convective zone.   

Our main objective is to take a wide class of profiles in the solar
interior, extracting from them the RSFP predictions for all neutrino
data \cite{SK}, \cite{Cl}, \cite{SAGE}, \cite{Gallex+GNO}, \cite{SNO1},
\cite{SNO2} available after the recently announced neutral current results 
from the SNO collaboration \cite{SNO2}  
and selecting those profiles which provide the best global 
fits based on a standard $\chi^2$ analysis. For these we evaluate the 
95\% and 99\% CL contours in the $\Delta m^2_{21}$, $B_0$ plane and 
obtain the corresponding predictions for the Borexino experiment 
\cite{Borexino}. We also consider fits to the rates separately. Our
calculation and fitting procedures are described in detail in our previous 
papers \cite{PA}, \cite{AP}, \cite{Astrop}. 

Our investigation proceeds along three main viewpoints: in the first of 
them we take the 'old' data set, i. e. the existing data prior to the 
announcement by the SNO collaboration of their new CC (reduced rate and 
day/night asymmetry) and NC results \cite{SNO2}. For the neutrino deuteron
cross section error values we use the result from a comparison between
refs. \cite{Kubodera}, \cite{Ying}. For the
Gallium rate we used the value ($74.7\pm5.13$) SNU \cite{Gallex+GNO}. 
In the second of these cases we include the SNO newly reported error 
values for the neutrino deuteron cross sections and CC reduced 
rate \cite{SNO2}. We also used the combined data from all Gallium experiments 
($72.4\pm4.7$) SNU. Thus the main feature of this second analysis type is a
reduction of all error bars, which necessarily results in an increase of the
$\chi^2$ in each case. In the third case we add the latest SNO results on CC 
including the day/night asymmetry and NC \cite{SNO2}. 

The paper is organised as follows: in section 2 we present our field
profiles in the radiative zone and core, and in the convective zone. We
consider seven profiles in the radiative zone and core. As for the
convective zone we take three profiles previously investigated in ref.
\cite{PA} and one in ref. \cite{Astrop}. We perform fits to the rates 
only and global fits in each of the three cases, selecting for the
global fits in the radiation zone and core those two which provide the 
best rate fits. We then select these two from the radiative zone and 
core and the best two from the convective zone which are used to determine 
the 95\% and 99\% CL contours around the best fit points. In section 3 we use 
these four contours to evaluate the corresponding Borexino predictions. Owing
to the characteristic shape of the RSFP survival probability, with a 
global minimum in the intermediate energy neutrino sector \cite{PA}, and
since Borexino is especially aimed at these neutrinos, the 
distinction between RSFP and oscillation scenarios will be quite possible
with Borexino \cite{AP1}. For two of 
the convective zone profiles such predictions were already obtained in 
ref. \cite{AP1} with the 'old' data set. The comparison between the 
Borexino predictions obtained with the 'old' and 'new' data sets presented
here shows that, while the central values remain practically unchanged, the 
ranges become substantially smaller, owing to the substantial decrease in 
the error bars. Consequently the possible distinction between the two 
scenarios (RSFP and oscillations) in Borexino will become even clearer 
with the present data than before \cite{AP1}. In section
4 we draw our main conclusions. We use throughout the BP'00 value for the
$^8B$ flux \cite{BP2000Bahc}. 


\section{RSFP Solutions}

\subsection{Solar Field Profiles}

We present in this subsection the solar field profiles used to obtain the
rate and global fits. We start with the core and radiation zone ones (see
also fig.1).

$Profile~RZ1$

\be
B=B_0[1-\left(\frac{x-x_c}{x_c}\right)^2]~~~~~|x| \leq x_c~,~x_c=0.16
\ee
\be
B=\frac{B_0}{cosh[8(x-x_c)]}~~x_c \leq |x| \leq x_m
\ee

$Profile~RZ2$

\be
B=B_0~exp\left(-\frac{x}{0.18}\right)~~~|x| \leq x_m.
\ee

$Profile~RZ3$

\be
B=B_0\left[1-\left(\frac{x}{x_m}\right)^2\right]~~~ |x| \leq x_m.
\ee

$Profile~RZ4$

\be
B=B_0\left(\frac{x}{x_c}\right)~~~~~~~~~|x| \leq x_c~,~x_c=0.356
\ee
\be
B=B_0\left[1-\frac{x-x_c}{x_m-x_c}\right]~~~x_c \leq |x| \leq x_m
\ee

$Profile~RZ5$

\be
B=B_0~~~~~~~|x| \leq x_c~,~x_c=0.188
\ee
\be
B=B_0~exp(-x)~~~~x_c \leq |x| \leq x_m
\ee

$Profile~RZ6$

\be
B=\frac{B_0}{cosh(6x)}~~~~~|x| \leq x_m
\ee

$Profile~RZ7$

\be
B=B_0\left(1-\frac{x}{x_m}\right)~~~~~|x| \leq x_m.
\ee

\begin{figure}[h]
\setlength{\unitlength}{1cm}
\begin{center}
\hspace*{-1.6cm}
\epsfig{file=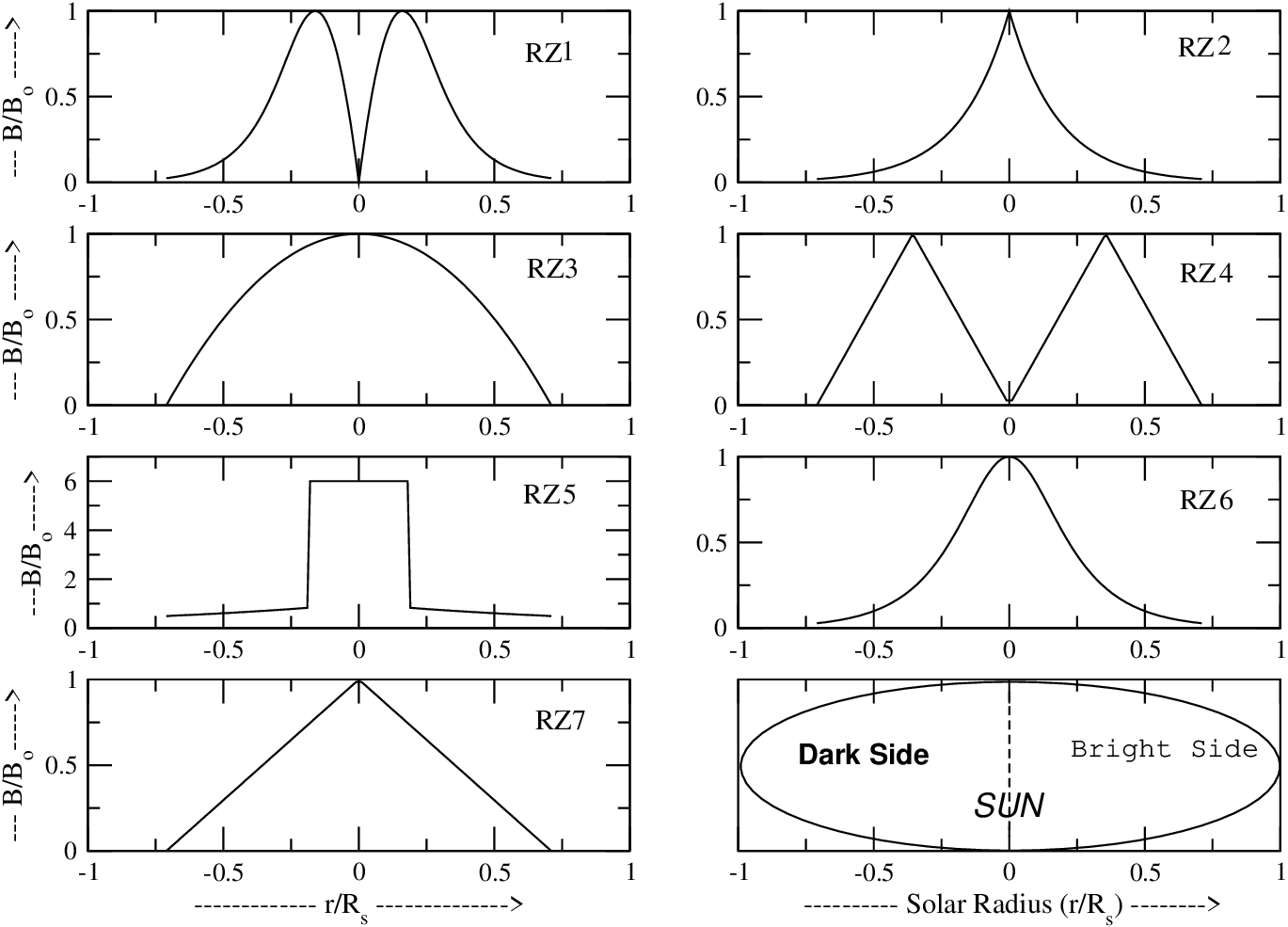,height=10.0cm,angle=0}
\end{center}
\vspace{-0.2cm}
{\small Figure 1: Solar field profiles in the radiative zone and core. RZ2
and RZ6 are the most favoured by the data.}
\end{figure}
In all cases $x$ is the fraction of the solar radius ($x=r/R_S$) and 
$x_m=0.713$, the bottom of the convective zone. All RZ profiles, defined
for ${x~^{>}}\!\!\!_{<}~0$, are taken to vanish for $|x| \geq x_m$.

\begin{figure}[h]
\setlength{\unitlength}{1cm}
\begin{center}
\hspace*{-1.6cm}
\epsfig{file=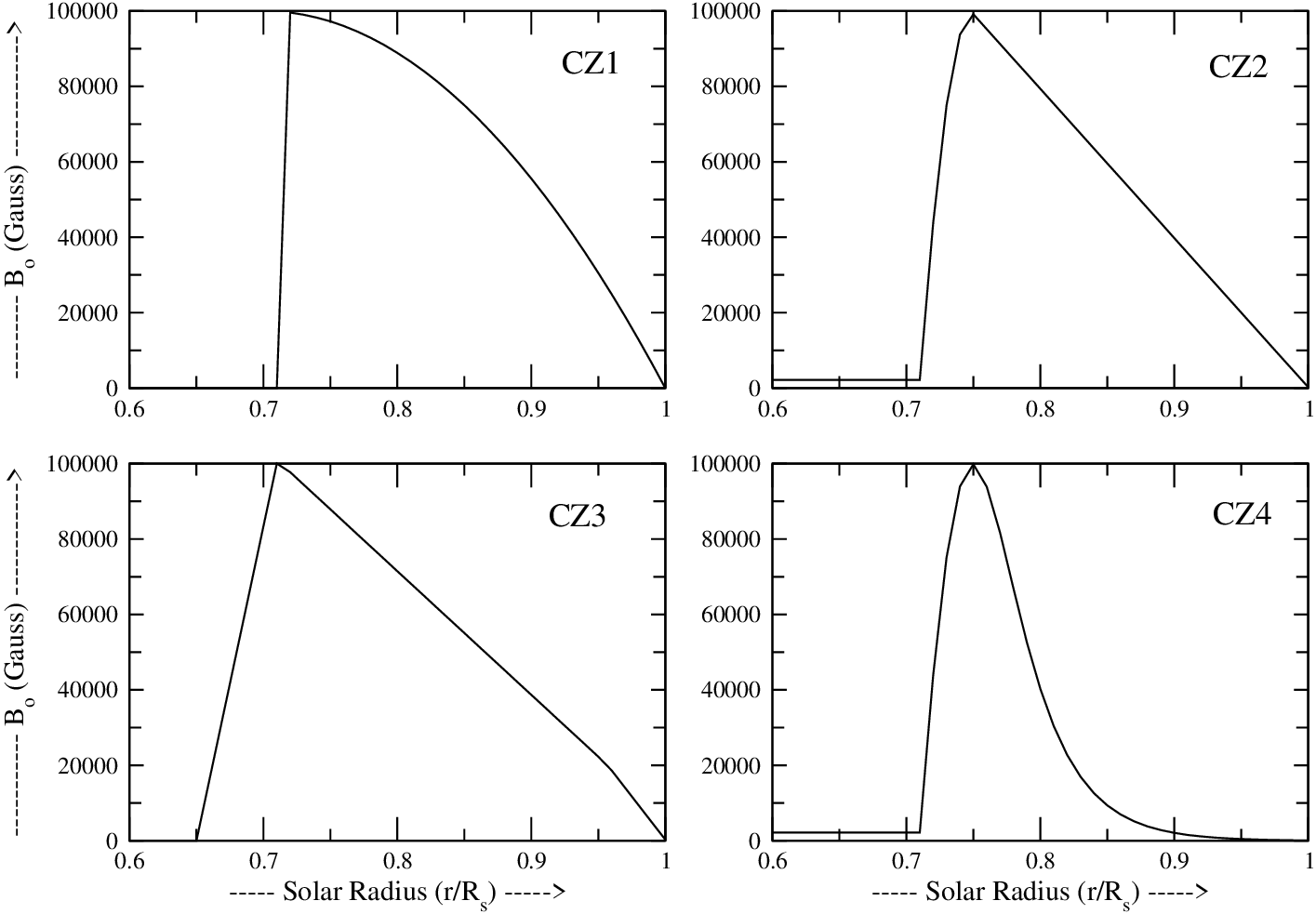,height=10.0cm,angle=0}
\end{center}
\vspace{-0.2cm}
{\small Figure 2: Solar field profiles in the convective zone. These are
selected from previous papers \cite{PA}, \cite{Astrop}, where they were
found to lead to the best fits in the light of the existing data at the time.}  
\end{figure}
For the convective zone, we take the following profiles (see also fig.2)

$Profile~CZ1$

\be
B=0~~~~~~~x \leq x_R~~~~x_R=0.71
\ee
\be
B=B_0\left[1-\left(\frac{x-0.7}{0.3}\right)^2\right]~~x \geq x_R
\ee

$Profile~CZ2$

\be
B=2.16\times10^3~~~,~~~x \leq 0.7105
\ee
\be
B=B_1\left[1-\left(\frac{x-0.75}{0.04}\right)^2\right]~,~0.7105<x<0.7483
\ee
\be
B=1.1494B_0[1-3.4412(x-0.71)]~,~0.7483 \leq x \leq 1
\ee

$Profile~CZ3$
\be
B=0~~,~~x\leq x_R~~~~~~~x_R=0.65
\ee
\be
B=B_0\frac{x-x_R}{x_C-x_R}~~,~~x_R\leq x \leq x_C~~x_C=0.713
\ee
\be
B=B_0+(x-x_C)\frac{2\times10^{4}-B_0}{0.957-x_C}~~,~~x_C\leq x\leq 0.957
\ee
\be
B=2.10^{4}+(x-0.957)\frac{300-2\times10^{4}}{1-0.957}~~,~~0.957\leq x\leq 1.
\ee

$Profile~CZ4$
\be
B=2.16\times10^3~~~,~~~x \leq 0.7105
\ee
\be
B=B_1\left[1-\left(\frac{x-0.75}{0.04}\right)^2\right]~,~0.7105<x<0.7483
\ee
\be
B=\frac{B_0}{cosh~30(x-0.7483)}~,~0.7483 \leq x \leq 1,~B_0=0.998B_1
\ee

Profiles CZ1, CZ2, CZ4 are respectively profiles 4, 7, and 6 of ref.\cite{PA}
and profile CZ3 is profile 4 of ref.\cite{Astrop}. Also profiles CZ3 and CZ4
were investigated in ref. \cite{AP1} as III and II and their Borexino 
predictions were then derived in the basis of the pre-SNO NC data. 

\begin{table}[h]
{\small Table I - Data from the solar neutrino experiments. Units are SNU for
Homestake and Gallium and $10^{6}cm^{-2}s^{-1}$ for SuperKamiokande and SNO.
See the main text for details.} 
\begin{center}
\begin{tabular}{lccc} \\ \hline \hline
Experiment &  Data      &   Theory   &   Data/Theory  \\ \hline
Homestake  &  $2.56\pm0.16\pm0.15$ & $7.7\pm^{1.3}_{1.1}$ & $0.332\pm0.05$  \\
Ga     &  $74.7\pm5.13$ & $129\pm ^8_6$ & $0.58\pm0.08$ \\
Ga (combined)    &  $72.4\pm4.7$ & $129\pm ^8_6$ & $0.56\pm0.07$ \\
SuperKamiokande&$2.32\pm0.085$ &
$5.05\pm^{1.0}_{0.7}$&$0.459\pm0.005\pm^{0.016}_{0.018}$\\
SNO CC (June 2001)&$1.75\pm{0.07}\pm^{0.12}_{0.11}\pm{0.05}$&$5.05\pm^{1.0}_{0.7}$
&$0.347\pm0.029$  \\ 
SNO CC (April 2002)&$1.76\pm{0.05}\pm{0.09}$&$5.05\pm^{1.0}_{0.7}$&$0.349\pm0.020$ \\ 
SNO CC ($A^{CC}_{D/N}$, April 2002)& $0.14\pm0.063\pm^{0.015}_{0.014}$ &$ 0 $ & $ 0 $ \\ 
SNO NC & $5.09\pm0.44\pm0.45$ & $5.05\pm^{1.0}_{0.7}$ & $ 1.01\pm0.13 $ \\ 
\hline
\end{tabular}
\end{center}
\end{table}


\subsection{Rate Fits}

The data on rates are summarized in table I and the RSFP best rate fits for all
11 profiles are presented in tables II, III. All fits including global ones 
(subsection 2.3) were obtained for the three analysis viewpoints mentioned in the 
introduction. Thus we consider separately:

(a) Four rates: Chlorine, SuperKamiokande, Ga and SNO CC total reduced rates. 
The CC total reduced rate from SNO and its error were taken from their first 
data announced in June 2001 while the errors for the neutrino deuteron cross 
sections were taken from a comparison between Kubodera's tables \cite{Kubodera} 
and the Paris potential results \cite{Ying}.

(b) Four rates: Chlorine, SuperKamiokande, Ga (combined) and SNO CC (new) total 
reduced rates. The CC total rate from SNO and its error as well as the deuteron 
cross section errors were taken from their data announced in April 2002. These are 
all substantially smaller than in case (a), resulting in an increase of the 
$\chi^2$ at the minima and, consequently, in a smaller spread for the Borexino 
prediction (see section 3), owing to the increased steepness of the $\chi^2$. 

(c) Six rates: Chlorine, SuperKamiokande, Ga (combined), SNO CC (new) total 
reduced rates, SNO CC day/night asymmetry ($A^{CC}_{D/N}$) and NC total reduced 
rate. All errors are as in case (b).  

\begin{table}[h]
{\small Table II: Rate fits for solar magnetic field profiles in the radiative 
zone (Eqs.(1)-(10)). For each profile the values of $\Delta m^2_{21}$ and $B_0$
are given at the best fit together with the corresponding $\chi^2$ and goodness
of fit (g.o.f.) for analysis cases (a), (b) and (c), respectively, described 
in the main text. The values of $\chi_{min}^2$ correspond to 2 d.o.f. (cases
(a), (b)) and 4 d.o.f (case (c)). See the text for more details.}
\begin{center}
\begin{tabular}{cccccc} \hline \hline
Profile & &$\Delta m^2_{21} (eV^2)$ & $B_{0}(G)$ &  $\chi^2_{rates}$  & g.o.f. \\
\hline
RZ1& (a)   & $2.87\cdot 10^{-6}$ & $6.79\cdot 10^{5}$ & 3.56 & 16.9 \\
&(b)   & $2.86\cdot 10^{-6}$ & $7.05\cdot 10^{5}$ & 5.85 & 5.4  \\
&(c)   & $2.86\cdot 10^{-6}$ & $6.86\cdot 10^{5}$ & 10.7 & 3.1  \\
\hline
RZ2&(a)   & $2.61\cdot 10^{-6}$ & $17.2\cdot 10^{5}$ & 1.47 & 48.0 \\
&(b)    & $2.55\cdot 10^{-6}$ & $17.4\cdot 10^{5}$ & 1.65 & 43.8 \\
&(c)    & $2.57\cdot 10^{-6}$ & $17.2\cdot 10^{5}$ & 6.44 & 16.9 \\
\hline
RZ3&(a)    & $6.85\cdot 10^{-6}$ & $2.8\cdot 10^{5}$ & 8.09 & 1.7 \\
&(b)    & $6.85\cdot 10^{-6}$ & $2.73\cdot 10^{5}$ & 11.1 & 0.4 \\
&(c)    & $6.85\cdot 10^{-6}$ & $2.95\cdot 10^{5}$ & 14.3 & 0.63 \\
\hline
RZ4&(a)    & $5.49\cdot 10^{-7}$ & $2.21\cdot 10^{5}$ & 9.81 & 0.6 \\
&(b)    & $5.54\cdot 10^{-7}$ & $2.27\cdot 10^{5}$ & 12.4 & 0.2 \\
&(c)    & $5.49\cdot 10^{-7}$ & $2.18\cdot 10^{5}$ & 18.7 & 0.1 \\
\hline
RZ5&(a)    & $4.35\cdot 10^{-6}$ & $2.61\cdot 10^{5}$ & 1.61 & 44.7 \\
&(b)    & $4.43\cdot 10^{-6}$ & $2.57\cdot 10^{5}$ & 3.00 & 22.3 \\
&(c)    & $4.04\cdot 10^{-6}$ & $2.56\cdot 10^{5}$ & 8.65 & 7.0 \\
\hline
RZ6&(a)    & $2.64\cdot 10^{-6}$ & $10.5\cdot 10^{5}$ & 1.64 & 44.1 \\
&(b)    & $2.60\cdot 10^{-6}$ & $10.7\cdot 10^{5}$ & 1.97 & 37.3 \\
&(c)    & $2.64\cdot 10^{-6}$ & $10.4\cdot 10^{5}$ & 6.81 & 14.6 \\
\hline
RZ7&(a)    & $7.36\cdot 10^{-7}$ & $4.01\cdot 10^{5}$ & 7.29 & 2.6 \\
&(b)    & $7.23\cdot 10^{-7}$ & $4.09\cdot 10^{5}$ & 7.68 & 2.2 \\
&(c)    & $7.78\cdot 10^{-7}$ & $3.92\cdot 10^{5}$ & 13.7 & 0.9 \\
\hline
\end{tabular}
\end{center}
\end{table}


In all cases (a), (b), (c) the free parameters in the analysis are the mass 
square difference between neutrino flavours $\Delta m^2_{21}$ and the peak
field value $B_0$. The typical ranges to be investigated are 
$10^{-7}eV^2\leq \Delta m^2_{21}\leq 5\times10^{-6}eV^2$, 
$0.1\times10^6G \leq B_0 \leq 2\times 10^6G$ for RZ profiles and 
$0.7\times10^{-8}eV^2\leq \Delta m^2_{21}\leq 3\times10^{-8}eV^2$, 
$0.1\times10^5G \leq B_0 \leq 1.5\times 10^5G$ for CZ ones.
With these choices the number of d.o.f. is 2 in cases (a)
and (b) and 4 in case (c). For profiles (RZ1-RZ7), in which case the magnetic
field extends over the neutrino production zone, we take for the survival 
probability the well known formula

\be
P=\frac{1}{2}+(\frac{1}{2}-P_C)cos2\theta_{i}cos2\theta_{0}
\ee
with the jump probability $P_C$ given by the Landau Zener approximation

\be
P_C=exp\left(-\pi\frac{2\mu^2B^2}{\Delta m^2_{21}/2E}0.09R_S\right)
\ee 
and which we integrate over the production regions and energy ranges for each
solar neutrino flux. In using this procedure, which avoids the numerical
integration of the neutrino evolution equations for each production bin,
care must be taken to account for those situations in which neutrinos are produced 
after the resonance, or the solar density is not large enough to ensure the
existence of a resonance and finally to account for the neutrinos which are 
produced in the far side of the sun. The production region and energy 
spectra were taken from \cite{BahcallHom}.  

For the convective zone profiles (CZ1-CZ4) the survival probabilities were 
obtained through the integration of the evolution equations as described in our 
previous work \cite{PA}.

The results of the 'rates only' analysis is shown in tables II for the radiative
zone and III for the convective zone profiles. Generically, it is seen that
the quality of the fits depends crucially on which data set is used. From
cases (a) to (c) the $\chi^2$ of the fits increases because the 
uncertainties improve in case (b) relative to (a) and because in
case (c) the 2.1${\sigma}$ day/night asymmetry of the CC event rate is taken
into account. This confronts the RSFP prediction of zero asymmetry. A
comparison between RZ and CZ profiles shows that the 'best' RZ profiles
produce fits of the same approximate quality as the CZ profiles. The latter
were chosen to be the 'best' from our previous experience \cite{PA}, \cite{Astrop}.
Hence it is seen that the data clearly shows no preference for 
a magnetic field either in the radiative or the convective zone. It is also
noteworthy that RZ profiles (table II) with the strongest field at the centre of 
the sun and a rapid decrease away from the centre are clearly favoured (RZ2, RZ6). 
The next best is RZ5 which clearly exhibits the same feature. We will select the 
best two (RZ2, RZ6) to perform the global fits described in the next subsection.

\begin{table}[h]
{\small Table III: Same as table II for the solar magnetic field profiles in
the convective zone (Eqs.(11)-(22)). The number of d.o.f. is 2 (cases (a), (b))
and 4 (case (c)).}
\begin{center}
\begin{tabular}{cccccc} \hline \hline
Profile & &$\Delta m^2_{21} (eV^2)$ & $B_{0}(G)$ &  $\chi^2_{rates}$  & g.o.f. \\
\hline
CZ1&(a)    & $1.27\cdot 10^{-8}$ & $9.7\cdot 10^{4}$ & 0.95 & 62.2 \\
& (b)    & $1.15\cdot 10^{-8}$ & $9.7\cdot 10^{4}$ & 1.17 & 55.7  \\
& (c)    & $1.16\cdot 10^{-8}$ & $9.68\cdot 10^{4}$ & 5.87 & 20.9  \\
\hline
CZ2&(a)    & $1.18\cdot 10^{-8}$ & $12.6\cdot 10^{4}$ & 1.17 & 55.7 \\
& (b)    & $1.09\cdot 10^{-8}$ & $12.6\cdot 10^{4}$ & 1.38 & 50.0  \\
& (c)    & $1.09\cdot 10^{-8}$ & $12.6\cdot 10^{4}$ & 6.05 & 19.5  \\
\hline
CZ3&(a)    & $1.42\cdot 10^{-8}$ & $9.8\cdot 10^{4}$ & 1.34 & 51.2 \\
& (b)    & $1.36\cdot 10^{-8}$ & $10.0\cdot 10^{4}$ & 2.14 & 34.3  \\
& (c)    & $1.37\cdot 10^{-8}$ & $9.92\cdot 10^{4}$ & 6.80 & 14.7  \\
\hline
CZ4&(a)    & $1.45\cdot 10^{-8}$ & $10.9\cdot 10^{4}$ & 1.72 & 42.3 \\
& (b)    & $1.43\cdot 10^{-8}$ & $11.2\cdot 10^{4}$ & 2.61 & 27.1  \\
& (c)    & $1.43\cdot 10^{-8}$ & $11.2\cdot 10^{4}$ & 7.39 & 11.7  \\
\hline
\end{tabular}
\end{center}
\end{table}

\subsection{Global Fits}

We selected from all 7 profiles in the radiative zone (RZ1-RZ7) those two which
provide the best rate fits and obtained the corresponding global fits. These
are RZ2, RZ6 (see tables IV and V). Global fits were performed for all four 
convective zone profiles. The global fit analysis follows 
viewpoints (a), (b) and (c) described in subsection 2.2 with the addition 
of the SuperKamiokande day/night spectrum for 1258 days \cite{SK} (19 day + 19
night energy bins) and the exclusion of the total SuperKamiokande rate. This 
exclusion avoids redundant information already present in the spectral bins 
and is common to most recent analyses. With these choices the number of d.o.f. 
is now in each case,

(a), (b): 3 rates + 38 spectral bins - 2 parameters = 39 d.o.f.

(c): 5 rates + 38 spectral bins - 2 parameters = 41 d.o.f.  

The global fits obtained for the selected profiles in the radiative zone and 
core are shown in table IV and those for the convective zone in table V. It
is seen that the quality of the global fits is the same for the best two 
profiles in the radiative zone (RZ2, RZ6) on one hand and all chosen four 
convective zone ones on the other (which were chosen as the best from our 
previous experience). All six exhibit a $\chi^2_{global}$ roughly equal or 
slighly smaller than the number of d.o.f. As in the case of the analysis of 
rates only, the radiative zone profiles that seem favoured by the data are 
those for which the field is the strongest at the solar centre with an almost
immediate rapid decrease away from the centre. 
Hence the favoured magnetic field profiles appear to satisfy a dipole structure
centered in the solar centre. Their shape also much resembles the solar matter 
density shape.

\begin{table}[h]
{\small Table IV: Global fits for solar magnetic field profiles in the
radiative zone (Eqs.(1)-(10)). The number of d.o.f. is 39 (cases (a), (b))
and 41 (case (c)).}
\begin{center}
\begin{tabular}{cccccc} \hline \hline
Profile & &$\Delta m^2_{21} (eV^2)$ & $B_{0}(G)$ & $\chi^2_{global}$ & g.o.f.\\
\hline
RZ2&(a)    & $2.61\cdot 10^{-6}$ & $16.3\cdot 10^{5}$ & 35.0 & 65.1 \\
& (b)    & $2.54\cdot 10^{-6}$ & $16.3\cdot 10^{5}$ & 35.0 & 65.1 \\
& (c)    & $2.54\cdot 10^{-6}$ & $16.7\cdot 10^{5}$ & 40.0 & 51.4 \\
\hline
RZ6&(a)    & $2.66\cdot 10^{-6}$ & $9.97\cdot 10^{5}$ & 35.0 & 65.5 \\
& (b)    & $2.59\cdot 10^{-6}$ & $9.99\cdot 10^{5}$ & 35.0 & 65.4 \\
& (c)    & $2.54\cdot 10^{-6}$ & $10.2\cdot 10^{5}$ & 40.1 & 51.2 \\
\hline
\end{tabular}
\end{center}
\end{table}

For the best profile in the radiative zone (RZ2) and the best one in the 
convective zone (CZ3) we show in the following the predictions at the 
best global fits for the experimentally measured quantities in cases (c) (all
new data). These are global fits 2 (c) in table IV and 3 (c) in table V which 
the reader can compare with the data given in table I. 

\newpage
$Profile~RZ2,~\chi^2_{gl}=40.01~(41~d.o.f.)$ 

$R_{Ga}=72.5~SNU,~R_{Cl}=2.64~SNU,~R_{SNO,CC}=0.354$ 

$A^{CC}_{D/N}=0,~R_{SNO,NC}/R^{st}_{SNO,CC}=0.968$

\vspace{0.5cm}
$Profile~CZ3,~\chi^2_{gl}=40.34~(41~d.o.f.)$ 

$R_{Ga}=72.5~SNU,~R_{Cl}=2.45~SNU,~R_{SNO,CC}=0.364$ 

$A^{CC}_{D/N}=0,~R_{SNO,NC}/R^{st}_{SNO,CC}=0.968$

\vspace{0.5cm}

All predictions are well within $1\sigma$ of the measured data, except for
the $A^{CC}_{D/N}$ asymmetry whose RSFP prediction is strictly zero.

Finally we choose the two radiative zone and the two convective zone
profiles providing the best global fits in case (c) to evaluate the
95\% CL ($1.96\sigma$) and 99\% CL ($2.58\sigma$) allowed regions in the 
$\Delta m^2_{21}$, $B_0$ plane. These global fits are 2 (c), 6(c) in 
table IV and 3 (c), 4 (c) in table V and the contours of the allowed 
regions are shown in fig.3. They are defined as the set of points satisfying
$\chi^2(\Delta m^2_{21},B_{0})-\chi^2_{min}=\Delta \chi^2(CL,~2~d.o.f.)$ with
$\Delta \chi^2(CL,~2~d.o.f.)=5.99, 9.21$ for 95\%CL and 99\% CL respectively. 
They will be used to evaluate the predictions for the Borexino 
experiment in the next section.
\begin{table}[h]
{\small Table V: Global fits for solar magnetic field profiles in the
convective zone (Eqs.(11)-(22)). The number of d.o.f. is 39 (cases (a),
(b)) and 41 (case (c)).}
\begin{center}
\begin{tabular}{cccccc} \hline \hline
Profile & &$\Delta m^2_{21} (eV^2)$ & $B_{0}(G)$ &  $\chi^2_{global}$  & g.o.f. \\
\hline
CZ1&(a)    & $1.25\cdot 10^{-8}$ & $9.54\cdot 10^{4}$ & 35.7 & 61.9 \\
& (b)    & $1.14\cdot 10^{-8}$ & $9.54\cdot 10^{4}$ & 35.7 & 62.1  \\
& (c)    & $1.11\cdot 10^{-8}$ & $9.60\cdot 10^{4}$ & 40.7 & 48.4  \\
\hline
CZ2&(a)    & $1.31\cdot 10^{-8}$ & $11.0\cdot 10^{4}$ & 36.1 & 60.1 \\
& (b)    & $1.22\cdot 10^{-8}$ & $11.0\cdot 10^{4}$ & 36.1 & 60.4  \\
& (c)    & $1.21\cdot 10^{-8}$ & $11.1\cdot 10^{4}$ & 41.1 & 46.6  \\
\hline
CZ3&(a)    & $1.25\cdot 10^{-8}$ & $9.54\cdot 10^{4}$ & 35.7 & 62.0 \\
& (b)    & $1.39\cdot 10^{-8}$ & $9.67\cdot 10^{4}$ & 35.4 & 63.6  \\
& (c)    & $1.38\cdot 10^{-8}$ & $9.80\cdot 10^{4}$ & 40.3 & 50.0  \\
\hline
CZ4&(a)    & $1.38\cdot 10^{-8}$ & $10.4\cdot 10^{4}$ & 35.5 & 63.1 \\
& (b)    & $1.38\cdot 10^{-8}$ & $10.5\cdot 10^{4}$ & 35.6 & 62.7  \\
& (c)    & $1.40\cdot 10^{-8}$ & $10.8\cdot 10^{4}$ & 40.7 & 48.4  \\
\hline
\end{tabular}
\end{center}
\end{table}

\begin{figure}[h]
\setlength{\unitlength}{1cm}
\begin{center}
\hspace*{-1.6cm}
\epsfig{file=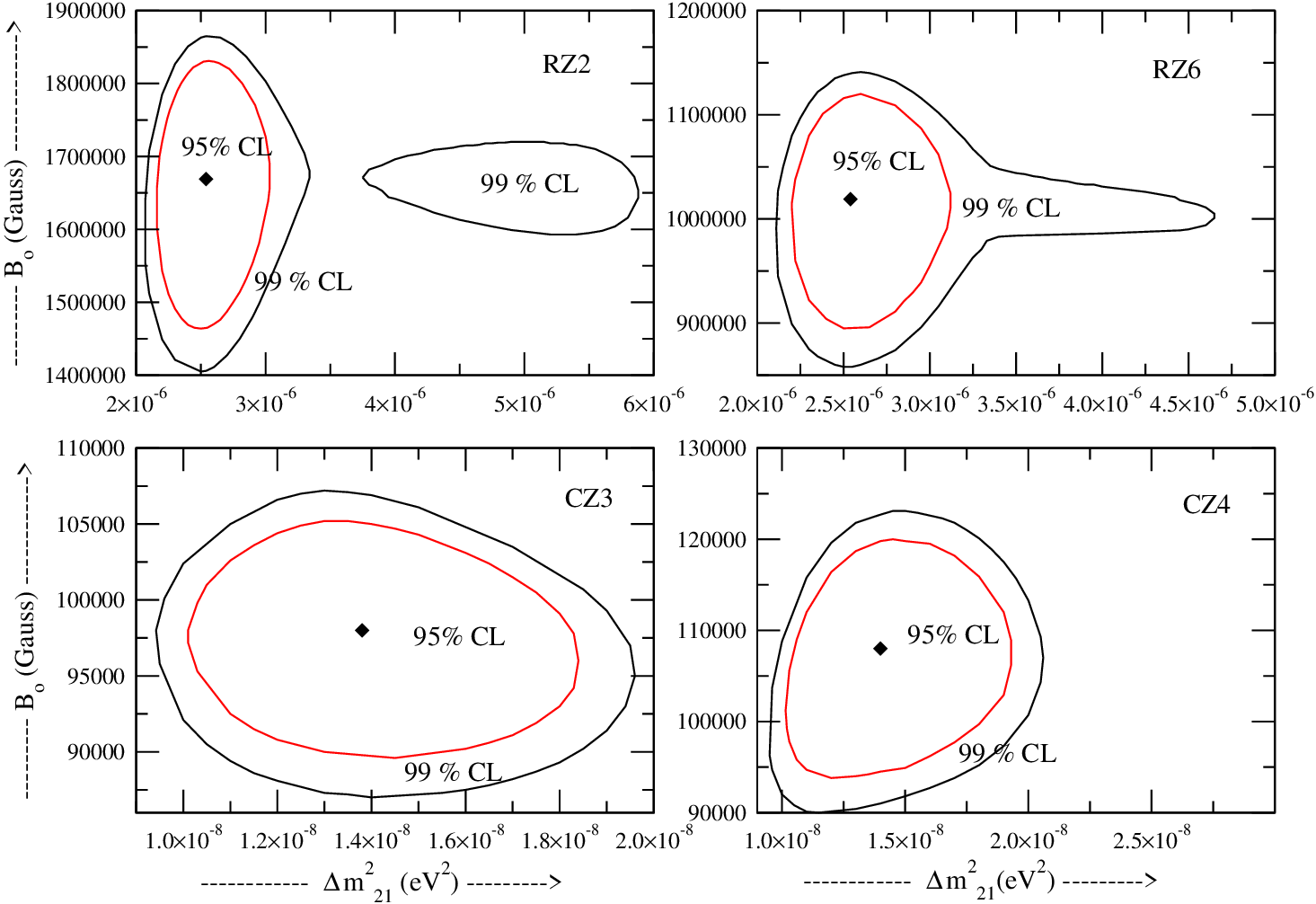,height=10.0cm,angle=0}
\end{center}
\vspace{-0.2cm}
{\small Figure 3: The 95\% and 99\%CL contours (for 2 d.o.f. $\Delta \chi^2=5.99, 
~9.21$ respectively) around the best global fits with analysis procedure (c) for
profiles RZ2, RZ6 in the radiative zone and core and CZ3, CZ4 in the convective 
zone of the sun. The best fit points are also shown. These correspond to $\chi^2=
40.0$ (profile RZ2), $\chi^2=40.1$ (profile RZ6), $\chi^2= 40.3$ (profile CZ3), 
$\chi^2=40.7$ (profile CZ4) with 41 d.o.f. See also tables IV and V.}
\end{figure}
\section{Predictions for Borexino}   

Except for the possible direct evidence that Kamland \cite{Kamland} may provide 
of the LMA solution, thus excluding RSFP as the dominant process for 
the solar neutrino deficit, no experiment other than Borexino is able to provide 
a positive distinction between oscillation solutions and RSFP. Hence it is
essential to investigate the 'best' RSFP predictions for Borexino. We chose
the 'best' two profiles in the radiative zone (RZ2, RZ6) and in the convective 
zone (CZ3, CZ4) and evaluated Borexino predictions at their best global fits, 
2 (c), 6 (c) (table IV) and 3 (c), 4 (c) (table V). We
also give in each case the upper and lower 95\% and 99\% CL
around the central values. These predictions, given as ratios between RSFP
event rates and standard event rates, and their confidence ranges are shown
in table VI. The analysis procedure, case (c), described in subsection 2.3,
involves all currently available solar neutrino data and best estimates
for the errors. 

Borexino predictions from RSFP were evaluated earlier \cite{AP1} for profiles 
CZ3, CZ4 (profiles III, II respectively in ref.\cite{AP1}) with the available
data as at December 2001. They were
$0.35\pm^{0.22}_{0.05}$ and $0.41\pm^{0.21}_{0.13}$ respectively for the
99\% CL. Comparing these with table VI it is seen that while the central values
hardly change, exhibiting a slight tendency for a decrease ($\simeq 4\%$), the
smaller errors from the neutrino deuteron cross sections and from the SNO and 
Ga rates lead to a sizeable reduction of the CL intervals. Such a reduction is
also observed in the oscillation predictions \cite{Beforeafter}. For RSFP
this is mainly
reflected in a decrease of the 95\% and 99\% CL upper limits, leading to the 
possibility of an even clearer distinction between RSFP and oscillation 
signatures in Borexino with the new data. In fact for the LMA solution such
a distinction is possible to more than $5.7\sigma$ for all four profiles
examined, whereas for the LOW solution all predictions are more than $4.5\sigma$ 
away (see table 2 of ref.\cite{Beforeafter} and our table VI). The only possible 
model dependence of RSFP predictions is contained in
the choice of the magnetic field profile, but this choice is severely constrained
by the requirement of fitting all solar data.
 
\begin{table}[h]
{\small Table VI: Predicted reduced event rates (rates assuming RSFP divided by
the standard solar model predictions) for Borexino using all 'new' data (case 
(c)).}
\begin{center}
\begin{tabular}{cccccc} \hline \hline
Profile  &  b.f. & min (95\%CL) & max (95\%CL) & min (99\%CL) & max (99\%CL) \\
\hline
RZ2      & 0.46  &   0.40       &    0.54     &  0.40        &  0.56  \\
RZ6      & 0.44  &   0.38       &    0.52     &  0.37        &  0.54  \\
CZ3      & 0.34  &   0.32       &    0.45     &  0.30        &  0.50  \\
CZ4      & 0.39  &   0.30       &    0.54     &  0.29        &  0.57  \\
\hline
\end{tabular}
\end{center}
\end{table}

\section{Summary and Conclusions}

Our main conclusions can be summarized in tables IV, V (cases labeled (c)), fig.3
and table VI. 

The objective of this paper is to present a statistical analysis of all
available solar neutrino data in the light of the RSFP solution to the solar
neutrino problem, after the recent presentation of the SNO neutral current results.
In addition to global fits and since these give, through the large number of  
spectral bins involved, a great significance to one single experiment, we also 
performed a separate analysis of rates only. Since the localization of the
strongest solar field is still unclear, we considered solar magnetic field
profiles both in the radiative zone and core, and in the convective zone of the sun.

The RSFP solutions do not predict any day/night effect, nor they imply any 
dependence of observable solar neutrino flux which follows the sunspot activity.   
Also in the basis of the Chlorine, SuperKamiokande and SNO experiments it will
be very hard to exclude RSFP solutions if the day/night asymmetries in SuperKamiokande 
and SNO remain consistent with zero. This difficulty is related to the fact that in 
the relevant solar neutrino energy ranges for these experiments, the survival
probability shape looks much the same for both RSFP \cite{PA} and the preferred 
oscillation solutions, LMA and LOW \cite{Beforeafter}. On the other hand, Gallium 
experiments will also be unable to tell the difference, in particular if only 
time averaged data are considered \cite{Sturrock}. Such a 'negative' situation is
however counterbalanced by the major difference in 
the above mentioned survival probability shapes in the intermediate energy neutrino
sector, mainly Be, at which the Borexino experiment is directly aimed. 
Hence the importance of Borexino predictions which we also investigated. Such
predictions were performed previously \cite{AP1} and they showed a clear distinction
between the two scenarios which, in the basis of the new data, has become better.
\footnote{There is a caveat here however due to the intrinsic error of the Borexino 
experiment \cite{deGouvea9904399}.} 

Altogether, radiative zone and core field profiles on one hand and convective
zone ones on the other are equally favoured by the data with fits of the same
quality as the best oscillation solution, the LMA one \cite{Beforeafter}. For
profiles in the radiative zone and core the data clearly prefer a strong field
at the centre of the sun with a rapid decrease thereafter. Interestingly enough
this shape of profiles follows a dipole structure centered at the solar
centre and closely resembles the density profile of the sun. 

Specific time signatures of the RSFP mechanism may be related with the possible
non-axially symmetric character of the solar field or the inclination of the
Earth's orbit. In the first case a time dependence would appear as a variation 
of the event rate with a period of 28 days, while in the second the possible
polar angle dependence of the solar field would cause a seasonal variation of the 
rate. Averaging rates over time erases all time dependent information that may be 
contained in the data. In fact a statistical analysis on the Gallium data performed 
by the Stanford group \cite{Sturrock} shows the existence of two peaks in the event
rates, which, while not providing conclusive evidence for RSFP, cannot be explained 
in the grounds of oscillations. It would be very important to independently repeat 
such analyses and to analyse the data in time bins in the future, especially if
Kamland shows a negative result.

\vspace{0.1cm}  
\noindent

{\em Acknowledgements.} We acknowledge useful discussions with Evgeni Akhmedov, 
Andr\'{e} de Gouv\^{e}a and correspondence with C. Pe\~{n}a-Garay. B.C.C. was 
supported by Funda\c{c}\~{a}o para a Ci\^{e}ncia e Tecnologia through Scolarship 
ref. SFRH/BPD/5719/2001.

\end{document}